\documentclass[11pt,twoside]{article}
\usepackage{asp2014}

\aspSuppressVolSlug
\resetcounters

\begin{document}

\title{Spreading the word - current status of VO tutorials and schools}

\author{Katharina~A.~Lutz$^1$, Mark Allen$^1$, Caroline Bot$^1$, Miriam Cort\'{e}s-Contreras$^2$, Sebastien Derriere$^1$, Markus Demleitner$^3$, Hendrik Heinl$^1$, Fran Jim\'{e}nez-Esteban$^2$, Marco Molinaro$^4$, Ada Nebot$^1$, Enrique Solano$^2$, and Mark Taylor$^5$}
\affil{$^1$CDS, Observatoire Astronomique de Strasbourg,  Universit\'{e} de Strasbourg, CNRS - UMR 7550, Strasbourg, France; \email{research@katha-lutz.de}}
\affil{$^2$Centro de Astrobiolog\'{i}a (INTA-CSIC), Madrid, Spain}
\affil{$^3$Astronomisches Rechen-Institut, Zentrum f\"{u}r Astronomie der Universit\"at Heidelberg, Heidelberg, Germany}
\affil{$^4$INAF - Osservatorio Astronomico di Trieste, Trieste, Italy}
\affil{$^5$H. H. Wills Physics Laboratory, University of Bristol, Bristol, United Kingdom}

\paperauthor{Katharina~A.~Lutz}{research@katha-lutz.de}{0000-0002-6616-7554}{CDS, CNRS, Observatoire Astronomique de Strasbourg}{}{Strasbourg}{}{67000}{France}
\paperauthor{Mark Allen}{}{0000-0003-2168-0087}{CDS, CNRS, Observatoire Astronomique de Strasbourg}{}{Strasbourg}{}{67000}{France}
\paperauthor{Caroline Bot}{}{}{CDS, CNRS, Observatoire Astronomique de Strasbourg}{}{Strasbourg}{}{67000}{France}
\paperauthor{Miriam Cort\'{e}s-Contreras}{}{}{Centro de Astrobiolog\'{i}a (INTA-CSIC)}{}{Madrid}{}{28850}{Spain}
\paperauthor{Sebastien Derriere}{}{}{CDS, CNRS, Observatoire Astronomique de Strasbourg}{}{Strasbourg}{}{67000}{France}
\paperauthor{Markus Demleitner}{}{}{Astronomisches Rechen-Institut, Zentrum f\"{u}r Astronomie der Universit\"at Heidelberg}{}{Heidelberg}{}{69120}{Germany}
\paperauthor{Hendrik Heinl}{}{}{CDS, CNRS, Observatoire Astronomique de Strasbourg}{}{Strasbourg}{}{67000}{France}
\paperauthor{Fran Jim\'{e}nez-Esteban}{}{}{Centro de Astrobiolog\'{i}a (INTA-CSIC)}{}{Madrid}{}{28850}{Spain}
\paperauthor{Marco Molinaro}{}{000-0001-5028-604}{INAF - Osservatorio Astronomico di Trieste}{}{Trieste}{}{34143}{Italy}
\paperauthor{Ada Nebot}{}{}{CDS, CNRS, Observatoire Astronomique de Strasbourg}{}{Strasbourg}{}{67000}{France}
\paperauthor{Enrique Solano}{}{}{Centro de Astrobiolog\'{i}a (INTA-CSIC)}{}{Madrid}{}{28850}{Spain}
\paperauthor{Mark Taylor}{}{}{H. H. Wills Physics Laboratory, University of Bristol}{}{Bristol}{}{}{United Kingdom}




\begin{abstract}

With some telescopes standing still, now more than ever simple access to archival data is vital for astronomers and they need to know how to go about it. Within European Virtual Observatory (VO) projects, such as {\bf AIDA (2008-2010)}, {\bf ICE (2010-2012)}, \textbf{CoSADIE (2013-2015)}, \textbf{ASTERICS (2015-2018)} and \textbf{ESCAPE (since 2019)}, we have been offering Virtual Observatory schools for many years. The aim of these schools are twofold: teaching (early career) researchers about the functionalities and possibilities within the Virtual Observatory and collecting feedback from the astronomical community. In addition to the VO schools on the European level, different national teams have also put effort into VO dissemination. The team at the Centre de Donn\'{e}es astronomiques de Strasbourg (CDS) started to explore more and new ways to interact with the community: a series of blog posts on AstroBetter.com or a lunch time session at the virtual EAS meeting 2020. The Spanish VO has conducted virtual VO schools. GAVO has supported online archive workshops and maintains their Virtual Observatory Text Treasures. In this paper, we present the different formats in more detail, and report on the resulting interaction with the community as well as the estimated reach.

\end{abstract}

\section{Introduction}

We report on recent European activities to inform about and train the astronomical community in the use of the Virtual Observatory (VO). The VO is an ever growing set of standards and protocols to unify the access to archives of astronomical data and to optimise the scientific exploitation of astronomical data. These standards and protocols are developed, discussed and defined by the International Virtual Observatory Alliance (IVOA). Within Europe, we have a long tradition of large, joint  projects, which aim at implementing these standards and protocols. One important aspect of these projects is the training of the astronomical community in using VO services. The current project is ESCAPE\footnote{https://projectescape.eu/}, which aims at implementing VO standards and protocols into the European Open Science Cloud. Within ESCAPE two "Science with interoperable data" Schools are planned. These schools have two goals: one goal is to expose participants to VO tools and services, so they can efficiently use them for their research. The second goal is to gather feedback and requirements for VO tools and services and the schools themselves from the participants. Besides the schools, we are also looking for other ways  to communicate about the VO and follow different avenues to interact with the community. This paper presents a short overview of our recent activities.

\section{Webpages and Blog Posts}

Making training material available online provides a sustainable long term resource for astronomers to tap into whenever needed. There are several of these resources, which we like to highlight here.

The EURO-VO webpage\footnote{http://www.euro-vo.org/} not only informs about ongoing work in European VO projects, but also hosts a collection of tutorials from previous VO schools. To ensure the sustainability of this webpage and improve the user experience, we have updated the content management system and design of the webpage. While designing, we put a particular focus on a clear design of the site presenting tutorials. These changes will go live soon after the submission of this proceedings article.

The Virtual Observatory Text Treasures (VOTT) is a formatted list of training material known to the VO registry, i.\,e. a central index.

With a large number of regular readers and several thousand followers on Twitter, AstroBetter\footnote{https://www.astrobetter.com/} is one of the most successful astronomy blogs. Between May and July 2020, we contributed four blog posts, which answered the question of how to access the VO-compliant services of the CDS with Python tools. The publication of each of the four articles was broadly announced by the AstroBetter team, which has provided great visibility. Since references to other resources have been given in these blog posts, they can serve as a first point of information regarding access to some VO services with Python.

\section{Virtual schools and training events}

EURO-VO Virtual Observatory schools and training events started as early as 2009 in the context of the  AIDA project (see also Tab.~2 in \citealp{Genova2015}). Throughout the ASTERICS project (2015-2019), we held four VO schools. With the current project, ESCAPE, in full swing we are now planning and organising the first ESCAPE "Science with interoperable data" school. Due to the current pandemic situation, this school is going to be virtual. One of the most important ingredients to the previous schools was the focus on usecases that participants brought along. These usecases were then tackled during the school using the newly learned techniques. Furthermore, tutors would support the students and help with the use-cases where necessary. We wish to continue this collaborative and hands-on approach in the upcoming, virtual school, which will take place in February 2021. In this event we will be able to accommodate up to 30 students. This school will contain six hands-on tutorials which will be presented in the mornings of 8 - 12th February 2021. The afternoons are reserved for students to work on the tutorials, while tutors will be available on Slack for questions. Students can then continue to work on their projects until 19th February 2021. On this day students and tutors will reconvene to present the results of the work on the usecases.

In addition to the VO schools of European projects, the Spanish Virtual Observatory\footnote{https://svo.cab.inta-csic.es} is regularly organising VO schools for the Spanish community. With the pandemic situation, these schools have also moved online. The first
remote SVO school has taken place in November 2020 as an extracurricular course for MSc students at the Universidad Complutense de Madrid. The platforms employed to carry out this school were zoom and a Slack workspace. The former was used for teaching and assessing the students during the hands-on sessions, and the latter was used as an off-line forum for asking questions and discussing other aspects of the VO and the school. The number of participants has been similar to previous editions (15-20). We did not observe a negative impact on the follow-up of the tutorials nor the communication with the participants. On the contrary, we noticed even more interaction between students and tutors compared to a face-to-face school, perhaps because we all attempted to supply the lack of the eye contact. The overall valuation rate has been very positive according to the participants and also from the tutors point of view. This challenging experience has taught us new ways of spreading knowledge and enriched ourselves for preparing the second remote SVO school, which will take place in December 2020.

The Shristi Astronomy course\footnote{https://shristiastro.com/} has been running throughout 2020 with many lectures. One of the lectures was presented by M. Taylor and focused on the very much VO-aware TOPCAT software packages. At the time of writing the recorded video of the live stream has been replayed almost 800\,times.

\section{Presence at virtual conferences and public talks}

Another place for interaction with the astronomical community are usually large conferences such as the IAU general assembly and the EAS and AAS scientific meetings. There we can not only present recent developments but also get feedback from the community. This year, the CDS presented their VO services in a dedicated Lunch Session at the virtual EAS scientific meeting. In addition to four talks, plenty of time was dedicated to questions and feedback from the community. Overall the session was well attended (> 80 attendees) with a lively discussion both in the Q\&A of the session and in the associated Slack channel.

Also at ADASS XXX (this conference) the contributions from the VO community, covering various topics from interoperability to tools and technologies, has been an important one, including 1 invited talk, 3 contributed ones, 1 tutorial, 2 BoF sessions and about a dozen poster contributions. Talks were both live streamed to a zoom session and on YouTube. Generally they would reach a relatively large audience of a couple hundred people during the conference. With the recordings of the talks being made available to the broader community afterwards, this number is likely to increase further. Posters were available to conference participants throughout the conference and were made publicly available after ADASS XXX finished. Generally, publicly available conference contributions such as recordings of ADASS XXX talks could also become a valuable source of information about the VO.

In addition to conferences, members of the European VO community also hold talks at various national events embracing open and FAIR data.

\section{Conclusion}
We are continuing to work on the dissemination of the Virtual Observatory. In the current, pandemic situation, we are developing new formats and investigate new outreach avenues. Our two main pillars are well maintained, accessible online resources, e.\,g. documents with tutorials, and interactive events, e.\,g. schools and sessions at conferences. First virtual interactive events have been very positive and we are looking forward to the upcoming virtual ESCAPE "Science with interoperable data" school.

\acknowledgements We acknowledge the support of ESCAPE (European Science Cluster of Astronomy and Particle Physics ESFRI Research Infrastructures) funded by the EU Horizon 2020 research and innovation program under the Grant Agreement n.824064.

\bibliography{P9-49}

\end{document}